\documentclass[sigconf]{acmart}

\usepackage{enumitem}
\usepackage{ulem}
\AtBeginDocument{%
  \providecommand\BibTeX{{%
    \normalfont B\kern-0.5em{\scshape i\kern-0.25em b}\kern-0.8em\TeX}}}

\copyrightyear{2022}
\acmYear{2022}
\setcopyright{acmcopyright}\acmConference[SIGIR '22]{Proceedings of the 45th International ACM SIGIR Conference on Research and Development in Information Retrieval}{July 11--15, 2022}{Madrid, Spain}
\acmBooktitle{Proceedings of the 45th International ACM SIGIR Conference on Research and Development in Information Retrieval (SIGIR '22), July 11--15, 2022, Madrid, Spain}
\acmPrice{15.00}
\acmDOI{10.1145/3477495.3532685}
\acmISBN{978-1-4503-8732-3/22/07}

\begin{document}

\title{Sequential/Session-based Recommendations: Challenges, \\Approaches, Applications and Opportunities}

\author{Shoujin Wang}
\affiliation{
  \institution{RMIT University, Macquarie University}
  \country{Australia}
  }

\author{Qi Zhang}
\email{qi.zhang-13@student.uts.edu.au}
\affiliation{%
  \institution{DeepBlue Academy of Sciences}
  \country{China}
}

\author{Liang Hu}
\email{rainmilk@gmail.com}
\affiliation{%
  \institution{Tongji University}
  \country{China}
}

\author{Xiuzhen Zhang}
\email{xiuzhen.zhang@rmit.edu.au}
\affiliation{%
 \institution{RMIT University}
 \country{Australia}
 }

\author{Yan Wang}
\email{yan.wang@mq.edu.au}
\affiliation{%
  \institution{Macquarie University}
  \country{Australia}
  }

\author{Charu Aggarwal}
\email{charu@us.ibm.com}
\affiliation{%
  \institution{IBM T. J. Watson Research Center}
  \country{US}
  }

\renewcommand{\shortauthors}{Wang, et al.}


\begin{abstract}
  In recent years, sequential recommender systems (SRSs) and session-based recommender systems (SBRSs) have emerged as a new paradigm of RSs to capture users’ short-term but dynamic preferences for enabling more timely and accurate recommendations. Although SRSs and SBRSs have been extensively studied, there are many inconsistencies in this area caused by the diverse descriptions, settings, assumptions and application domains. 
  There is no work to provide a unified framework and problem statement to remove the commonly existing and various inconsistencies in the area of SR/SBR. There is a lack of work to provide a comprehensive and systematic demonstration of the data characteristics, key challenges, most representative and state-of-the-art approaches, typical real-world applications and important future research directions in the area. This work aims to fill in these gaps so as to facilitate further research in this exciting and vibrant area.
\end{abstract}





\maketitle

\section{Introduction}
Recommender systems (RSs) have been playing an increasingly important role in informed consumption, and decision-making in the current era of information explosion and digitized economy~\cite{wang2021survey}. In recent years, sequential recommender systems (SRSs) and session-based recommender systems (SBRSs) have emerged as a new paradigm of RSs to capture users' short-term but dynamic preferences for enabling more timely and accurate recommendations~\cite{wang2019sequential}. SRSs and SBRSs have been quite important and popular research areas in the recommendation communities, which have attracted much attention from both academia and industry. SRSs and SBRSs are highly correlated and similar in terms of the input, output and recommendation mechanism, and most of the representative approaches for building SRSs and SBRSs are very similar. Therefore, we present this work to cover both SRSs and SBRSs. 

The key challenge of building SRSs/SBRSs lies in how to comprehensively learn the complex dependencies embedded within and between sequences/sessions to accurately infer users' timely and dynamic preferences~\cite{wang2021survey}. In recent years, there has been some promising progress in tackling this challenge, including, e.g., Markov chain based approaches~\cite{rendle2010factorizing}, distributed representation based approaches~\cite{wang2018attention}, recurrent neural network (RNN) based approaches~\cite{hidasi2015session}, graph neural network (GNN) based approaches~\cite{wu2019session,wang2021graph}, reinforcement learning-based approaches~\cite{zhao2017deep} and contrastive learn-ing-based approaches~\cite{qiu2022contrastive}.  

Although SRSs and SBRSs have been extensively studied in recent years, there are many inconsistencies within each area and/or between both areas, caused by the diverse description terms, scenario settings, employed assumptions and application domains. There is a lack of a unified framework to well categorize them, and there are no unified problem statements for the research problem(s)~\cite{wang2021survey}. A few tutorials have focused on sequence-aware recommender systems~\cite{Massimowww19}, deep learning-based sequential recommendations~\cite{FangICWE19}, and session-based recommendation on GPU~\cite{Moreirarecsys21}. However, there is no work to provide a unified framework and problem statement to remove the commonly existing and various inconsistencies in the areas of SRSs and SBRSs. There is a lack of work to provide a comprehensive and systematic demonstration of the data characteristics, key challenges, most representative and state-of-the-art approaches, typical real-world applications and important future research directions in the area of SRSs and SBRSs. This work aims to fill in these gaps so as to facilitate further research in this exciting and vibrant area.

\begin{figure}[!t]
	\centering
	\includegraphics[width=.47\textwidth]
	{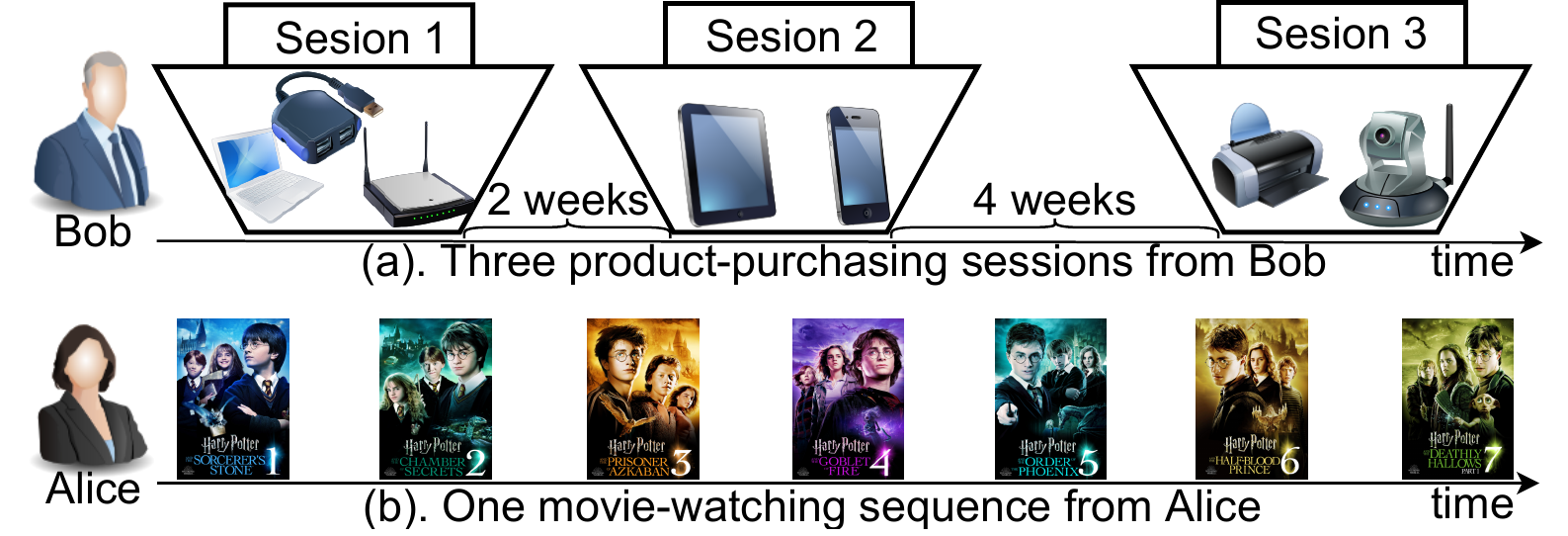}
	\vspace{-1em}
	\caption{Session data vs. sequence data, from ~\cite{wang2021survey}}
	\label{fig_sequence}
	\vspace{-2em}
\end{figure} 

\section{Related Work}
There are some surveys and tutorials focusing on the topic of SRSs or SBRSs. For SRSs, Quadrana et al. performed a comprehensive survey~\cite{quadrana2018sequence} together with two tutorials~\cite{Massimowww19,Massimorecsys18} at WWW 2019 and RecSys 2018 on sequence-aware recommender systems, which talk about the recommendation task, algorithms and evaluations of SRSs. Fang et al. provided a survey~\cite{fang2019deep} and presented a tutorial~\cite{FangICWE19} at ICWE 2019 both on deep learning based sequential recommendations. They discussed various aspects of SRSs including the concepts, algorithms, inﬂuential factors, and evaluations; Wang et al.~\cite{wang2019sequential} conducted a brief review on the challenges, progress and prospects of SRSs. Regarding SBRSs, to the best of our knowledge, there is only one comprehensive survey on SBRSs~\cite{wang2021survey} to systematically discuss the session-based recommendation problem, data characteristics, recent progress, approach taxonomy, applications and future directions. Ludewig et al.~\cite{ludewig2021empirical} conducted an empirical study on some representative SBRS algorithms while Gabriel et al.~\cite{Moreirarecsys21} provided a tutorial on session-based recommendation on GPU. 

These existing works have great value in treating specifics of research in more detail in the area of SRSs or SBRSs. However, they often focus on either SRSs or SBRSs, and none of them can systematically focus on both SRSs and SBRSs to systematically talk about the difference and similarities of SRSs and SBRSs, as well as to address the commonly existing inconsistencies w.r.t. concepts, settings, etc. between them. This work is well complementary to those related works by providing a more complete summarization of sequential and session-based recommendations with an emphasis on the problem statement, data characteristics and challenges, applications and prospects, and comprehensive analysis of all kinds of state-of-the-art approaches, models and algorithms. Specifically, it performs a comprehensive review of the latest survey papers on the sequential and session-based recommendations.

\begin{figure*}[!t]
	\centering
	\includegraphics[width=.86\textwidth]
	{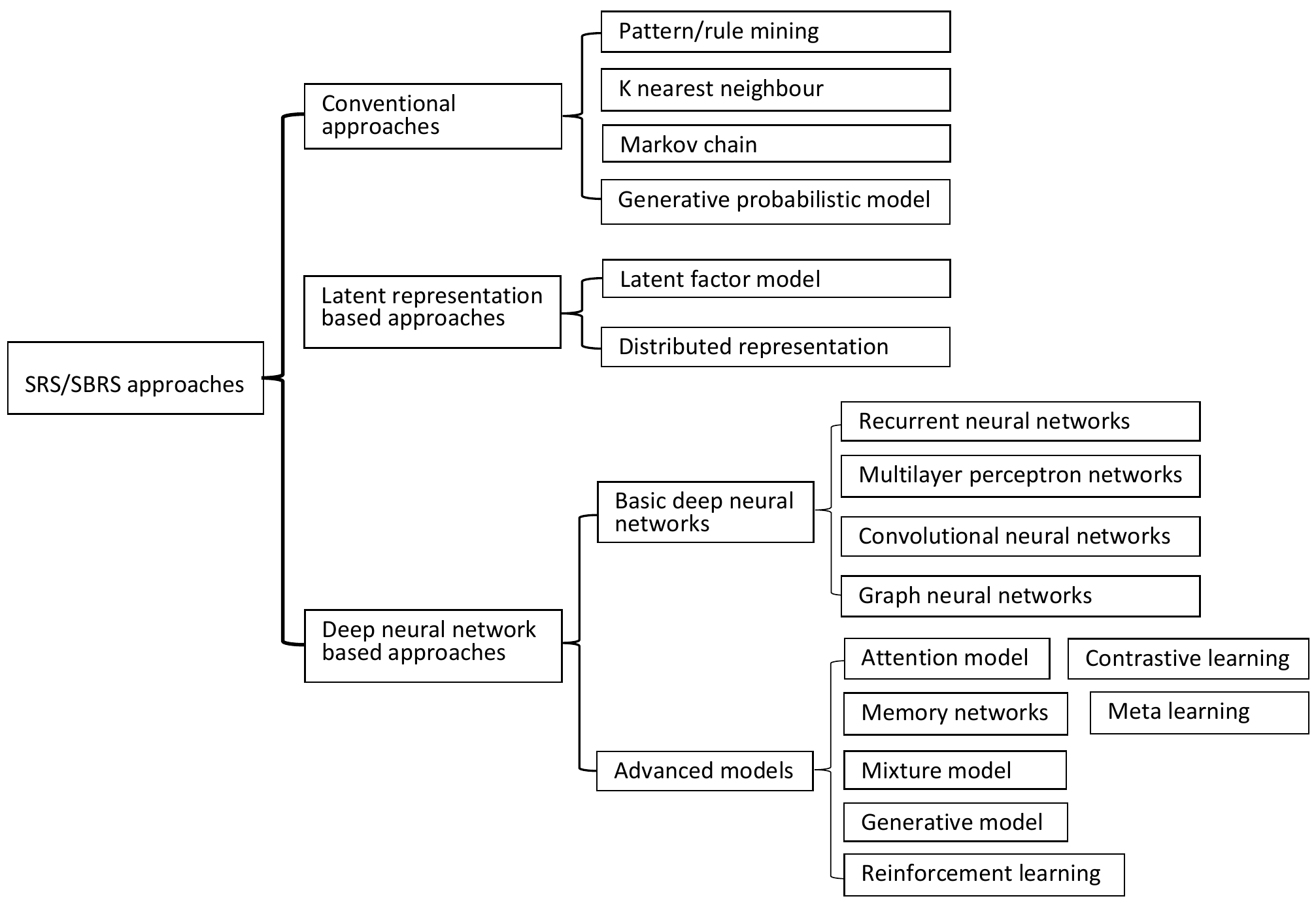}
	\vspace{-1em}
	\caption{The categorization of SBRS approaches, referred to \cite{wang2019sequential,wang2021survey}}
	\label{fig_class}
	\vspace{-0.5em}
\end{figure*}

\section{An Overview of This Work}
This work will perform a systematic and high-level review of the most notable works to date on SRSs/SBRSs. It will contain five parts: 
\begin{itemize}[leftmargin=*]
    \item{Part 1 Introduction and Problem Statement. This part will first introduce the background of SRSs and SBRSs with an emphasis on the comparison between them, followed by a unified problem statement of SRSs/SBRSs.}
    \item{Part 2 Data Characteristics and Challenges. This part will thoroughly analyze the characteristics of data used for SRSs and SBRSs and the main challenges triggered by them.}
    \item{Part 3 Sequential/Session-Based Recommendation Approaches. This part will first provide a classification scheme to well organize all the existing approaches to SRSs and SBRSs and then highlight the most recent advance in each class of approaches.}
    \item{Part 4 Applications and Algorithms. This part will introduce both the traditional and emerging real-world applications of SRSs and SBRSs and a collection of representative and state-of-the-art SRSs/SBRSs algorithms together with public datasets.}
    \item{Part 5 Future Opportunities. This part will discuss some of the most promising directions in the area and conclude this tutorial.}
 \end{itemize}

\section{Background and Problem Statement}

\subsection{SRSs vs SBRSs}\label{SRSvsSBRS}
Generally, SRSs and SBRSs take sequence data and session data as its input respectively. A session is a set of interactions with clear boundary and the interactions may be ordered or unordered. A sequence is a list a elements (such as item IDs) with clear chronological order. SBRSs either predict the next interaction(s) based on the given historical interactions within a session, or predict the future session (e.g., the next-basket) based on the historical sessions, which mainly depends on the intra- or inter-session dependencies. In comparison, SRSs predict the following elements of a sequence given the historical elements in the sequence, which mainly relies on the sequential or temporal dependencies over the elements inside each sequence~\cite{wang2021survey,wang2019learning}.

\subsection{Sequential/Session-based Recommendation Problem Statement}
\label{sec:problem}
There are five entities (namely User, Item, Action, Interaction and Sequence/Session) involved in sequential/session-based recommendation scenarios and they constitute the foundations for defining the sequential/session-based recommendation research problems. We first provide a brief introduction of each of them and then define the sequential/session-based recommendation research problem. 
\begin{itemize}[leftmargin=*]

\item{User and User Properties.}

\item{Item and Item Properties.}

\item{Action and Action Properties.} Actions refers to users' actions on items, such as clicks, views, purchases.

\item{Interaction and Interaction Properties.} An interaction is a triplet of $\left \langle user, action, item \right \rangle$. 

\item{Sequence/Session and Their Properties.} A sequence or a session is a set of interactions and can be characterised by a set of properties including sequence/session length, internal order, action type (purchase or click, etc.), user information availability and data structure. A detailed illustration was provided by Wang et al.~\cite{wang2021survey}.
\end{itemize}

\subsubsection*{Problem Statement}
Sequential/session-based recommendation is often formalized as a next-item or next-basket prediction problem. Specifically, given a user's historical interaction information, an RS is built to predict the user's future interactions, such as her/his next item, next basket to be purchased, or next POI to be visited. The main work mechanism is to first accurately learn the complex dependencies embedded in users' interaction behaviours and then employ the learned dependencies as a signal to guide the following prediction. So the main challenge/task lies in dependency learning.  

\section{Data Characteristics and Challenges}
\label{sec:datachar}

This section will introduce the unique characteristics of data used for SRSs/SBRSs and the special challenges they brought to SRSs/SBRSs from five dimensions: (1) sequence/session length, (2) the internal order within sequences/sessions, (3) the type of actions within sequences/sessions, (4) user information, and (5) sequence/session data structure. The outline of this section is listed below. 

\begin{itemize}[leftmargin=*]

\item{Characteristics and Challenges Related to Sequence/Session Length.} According to length, a sequence/session can be classified into long or short sequence/session, which has different challenges. For instance, the challenges for long sequence/sessions lie in how to learn long-range and/or high-order dependencies while that for short sessions lies in how to learn enough dependency information with limited interactions.  

\item{Characteristics and Challenges Related to Internal Order.} For sequences, there is clear order within each sequence, while for sessions, they may be ordered, unordered and flexibly ordered sessions. Sessions with different types of orders often have different challenges when making recommendations based on them.    

\item{Characteristics and Challenges Related to Action Type.} The action can be purchase, click, view, and add to cart. So a sequence/session can be based on each of them or a combination of any of them, leading to single-action-type sequence/sessions and multi-action-type sequence/sessions. The main challenges for multi-action-type sequence/sessions lie in how to effectively learn the intra- and inter-action type dependencies.  

\item{Characteristics and Challenges Related to User Information.}
Sequences/sessions can be divided into anonymous ones and non-anonymous ones. Usually, anonymous ones can be more challenging for dependency learning.  

\item{Characteristics and Challenges Related to Data Structure.} Session data can be divided into single-level sessions and multi-level sessions according to the number of structure levels involved~\cite{wang2021survey}. 

\end{itemize}

\section{SRS/SBRS Approaches}
\label{sec:taxonomy}

This section first describes the classification taxonomy of SRS/SBRS approaches, and then comprehensively compare them.    

\subsection{A Classification of Approaches}
As shown in Figure 1, the approaches are classified into three classes: (1) conventional approaches, (2) latent representation-based approaches, and (3) deep neural network-based approaches. The first class includes four sub-classes: pattern/rule mining-based approaches, K nearest neighbor approaches,  Markov chain approaches, and generative probabilistic approaches. The second class includes two sub-classes: latent factor model-based approaches and distributed representation-based approaches. The third class includes two sub-classes: basic deep neural network (i.e., RNN, MLP, CNN, GNN) based approaches and advanced model (i.e., attention mechanism, memory network, mixture model, generative model, reinforcement learning, contrastive learning, meta learning) based approaches~\cite{wang2021survey,wang2019sequential}. 

\subsection{Conventional Approaches}
\label{sec:conv_approaches}
Conventional approaches utilize conventional data mining or machine learning approaches to build SRSs/SBRSs \cite{wang2021survey}. This section will introduce four classes of conventional approaches for SRSs/SBRSs, and then compare them. The outline of this part is listed below.  
\begin{itemize}[leftmargin=*]

\item{Pattern/Rule Mining based SRSs/SBRSss.}

\item{K Nearest Neighbour based Approaches.}

\item{Markov Chain based Approaches.}

\item{Generative Probabilistic Model based Approaches.}
\end{itemize}

\subsection{Latent Representation Approaches for SRSs/SBRSs}
\label{sec:latent_rep}
Latent representation approaches first learn a low-dimensional latent representation for
each interaction from sequences or sessions (usually with shallow models) and then employ the learned representation as to the input of the subsequent recommendation task. This section will first introduce two classes of latent representation approaches, followed by a comparison between them. 
\begin{itemize}[leftmargin=*]
\item{Latent Factor Model based Approaches.}

\item{Distributed Representation based Approaches.}
\end{itemize}


\subsection{Deep Neural Network Approaches for SRSs/SBRSs}
\label{sec:dnn_approaches}

Deep neural network approaches mainly take advantage of the powerful capabilities of deep neural models in learning the complex dependencies within or between sequences/sessions for recommendations \cite{wang2021survey}, and they can be roughly classified into basic approaches and advanced approaches. Each basic deep neural approach is built on a basic deep neural network (e.g., RNN) while each advanced approach is based on one or more advanced neural models (attention model). This section will first introduce the four classes of basic approaches and five classes of advanced approaches and then compare them. The outline of this section is listed below.   
\subsubsection{Basic Deep Neural Network based Approaches}

\begin{itemize}[leftmargin=*]
    \item{Recurrent Neural Networks (RNN) based Approaches}
    \item{MultiLayer Perceptron (MLP) networks based Approaches}
    \item{Convolutional Neural Networks (CNN) based Approaches}
    \item{Graph Neural Networks (GNN) based Approaches}
\end{itemize}

\subsubsection{Advanced Model based Approaches}

\begin{itemize}[leftmargin=*]
\item{Attention Model based Approaches}

\item{Memory Networks based Approaches}

\item{Mixture Model based Approaches}

\item{Generative Model based Approaches}

\item{Reinforcement Learning (RL) based Approaches}

\item{Contrastive Learning based Approaches}

\item{Meta Learning based Approaches}
\end{itemize}


\subsection{A Comparison of Diﬀerent Classes of Approaches}
This part will provide comparisons between different classes of approaches from multiple perspectives, including the work mechanism of each class, the learned dependency types, and the research trend. Due to the limited space, please refer to \cite{wang2019sequential,wang2021survey} for detailed comparisons.

\section{SRS/SBRS Applications and Algorithms}
\label{sec:app_alg_datasets}
This section will first demonstrate the real-world applications of SRSs and SBRSs, including both the applications in conventional domains and emerging domains~\cite{wang2021survey}. Then, we will summarize 24 representative and state-of-the-art open-source algorithms for building SRSs and SBRSs and 13 commonly used public real-world datasets for testing the performance of these algorithms, the details of them can be found in ~\cite{wang2021survey}. 

\subsection{Applications}

\textbf{Conventional applications:}
\begin{itemize}[leftmargin=*]
\item{E-commerce domain: Next-item/basket recommendation.}
\item{Media, entertainment domain: Next news/web-page/song/movie
/video recommendation.}
\item{Tourism domain: Next-POI recommendation.}
\end{itemize}

\noindent\textbf{Emerging applications:}
\begin{itemize}[leftmargin=*]
\item{Finance domain: Next-trading/investment recommendation.}
\item{Healthcare domain: Next-treatment/medicine recommendation.}
\end{itemize}

\section{Prospects and Future Directions}
\label{sec:prospects_future}
This section will outline the following eight promising prospective research directions in the areas of SRSs and SBRSs.  
\begin{itemize}[leftmargin=*]
\item{SRSs/SBRSs with General User Preference}

\item{SRSs/SBRSs Considering More Contextual Factors}

\item{SRSs/SBRSs with Cross-domain Information}

\item{SRSs/SBRSs by Considering More User Behaviour Patterns}

\item{SRSs/SBRSs with Constraints}

\item{Interactive SRSs/SBRSs}

\item{Online or Streaming SRSs/SBRSs}

\item{Trustworthy and Responsible SRSs/SBRSs}

\end{itemize}

\begin{acks}
This work was supported by ARC Discovery Project DP200101441.
\end{acks}

\normalem
\bibliographystyle{ACM-Reference-Format}
\bibliography{Ref}

\end{document}